\documentclass[
 aip,
 jmp,%
 amsmath,amssymb,
 reprint,%
]{revtex4-1}

\usepackage{graphicx}
\usepackage{dcolumn}
\usepackage{bm}

\begin{document}

\preprint{}

\title{Graphene Transfer with Reduced Residue}



\author{Michael Her}
\affiliation{Institute of Optics, University of Rochester, Rochester, NY 14627, USA}

\author{Ryan Beams}
\affiliation{Institute of Optics, University of Rochester, Rochester, NY 14627, USA}

\author{Lukas Novotny}
\affiliation{Institute of Optics, University of Rochester, Rochester, NY 14627, USA}
\affiliation{ETH Z{\"u}rich, Photonics Laboratory, 8092 Z{\"u}rich, Switzerland}


\date{\today}

\begin{abstract}
We present a new transfer procedure for graphene using acetic acid, which removes the residue that is common in standard acetone treatments.  Post-transfer samples cleaned with acetic acid and acetone were characterized using Raman spectroscopy and atomic force microscopy for comparison.  We further illustrate the quality of our transfer process by using fluorescence quenching to create an optical map of surface contaminants.
\end{abstract}


\maketitle 


Graphene has demonstrated promise for future semiconductor and photonic technologies due to its unique electrical and optical properties, including an unusual quantum Hall effect combined with high carrier mobility~\cite{zhang05,novoselov07,bolotin09,du09}, universal optical conductivity~\cite{nair08,kuzmenko08,mak08}, and linear electronic dispersion near the neutrality point~\cite{bonaccorso10,geim07,neto09}.  A possible route for many applications, such as photovoltaic devices, organic LEDs, photodetectors, touch screens, and flexible smart windows, is the precise deposition of graphene on processed substrates~\cite{bonaccorso10}.  Thus far, considerable effort has been invested into procedures to transfer graphene between substrates~\cite{reina08,li09b,liang11,goossens12,bonaccorso10}.  In the standard chemical transfer procedure, a graphene flake is transferred to the target substrate using Poly(methyl methacrylate) (PMMA).  The PMMA is dissolved using acetone, leaving the graphene flake on the desired substrate\cite{reina08,li09b}.  While this is a straightforward procedure, the acetone treatment frequently fails to fully remove the PMMA, leaving a residue on the graphene surface and the substrate\cite{liang11}.  Several techniques have been demonstrated to remove this residue, including using a modified RCA cleaning process and mechanically sweeping away the contamination using an atomic force microscope (AFM)\cite{liang11,goossens12}.  However, these techniques involve either complicated wet chemistry or are limited to cleaning only a local area.\\[-1.2ex]

\begin{figure}[b!]
\begin{center}
\vspace{-2ex}\includegraphics[width = 8.1cm]{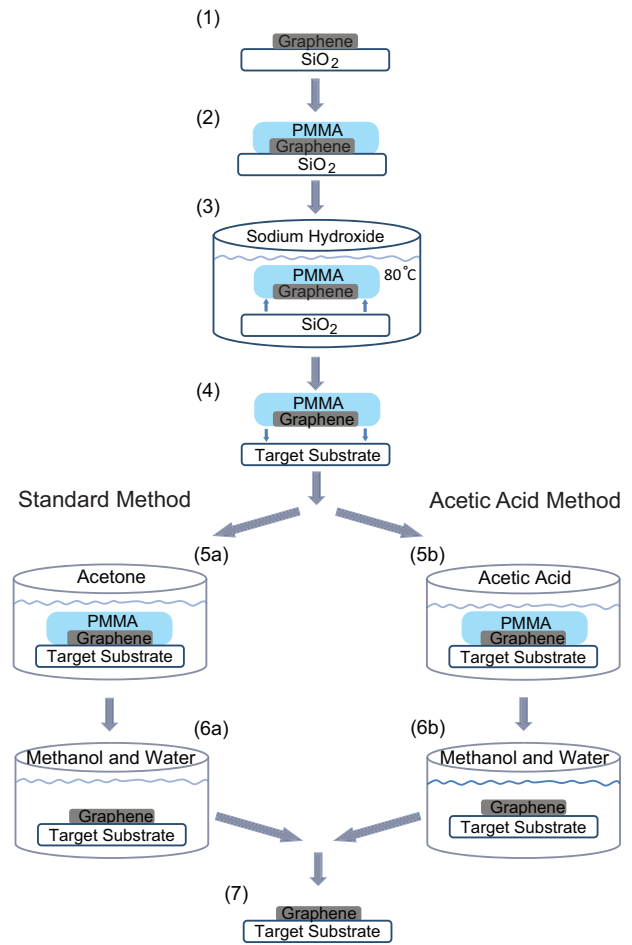}%
\caption{Illustration of graphene transfer.  PMMA is deposited on a graphene layer and cured at 115\,$^\circ$C for 2\,min.  The graphene-PMMA stack is then detached from the substrate in a sodium hydroxide bath and subsequently deposited on a target substrate. Finally, the PMMA is dissolved using acetone or acetic acid and then rinsed in a mixture of methanol and water.}
\label{trans-proc}
\end{center}
\end{figure}

In this paper, we present an alternative graphene transfer process.  Our process is similar to the transfer procedure introduced in Ref. \cite{reina08}.  This procedure starts by coating the substrate and graphene flake with PMMA.  By using a sodium hydroxide solution, the graphene-PMMA stack is lifted from the initial substrate and placed onto the target substrate.   Finally, the PMMA is dissolved using acetone.  However in our procedure, after transfer to the target substrate, the graphene-PMMA stack is placed in acetic acid for 24\,hours.  Afterwards, the sample is cleaned in methanol.  Both transfer procedures are illustrated in Figure $\,$\ref{trans-proc}.  Samples were prepared using both the standard acetone procedure and our acetic acid method.  Subsequently the two samples have been compared using differential optical contrast microscopy (DIC), Raman spectroscopy, and AFM.  Finally, we used fluorescence quenching to further illustrate the improvement in the quality of the acetic acid method.  \\[-1.2ex]

\begin{figure}[t!]
\begin{center}
\includegraphics[width = 8.5cm]{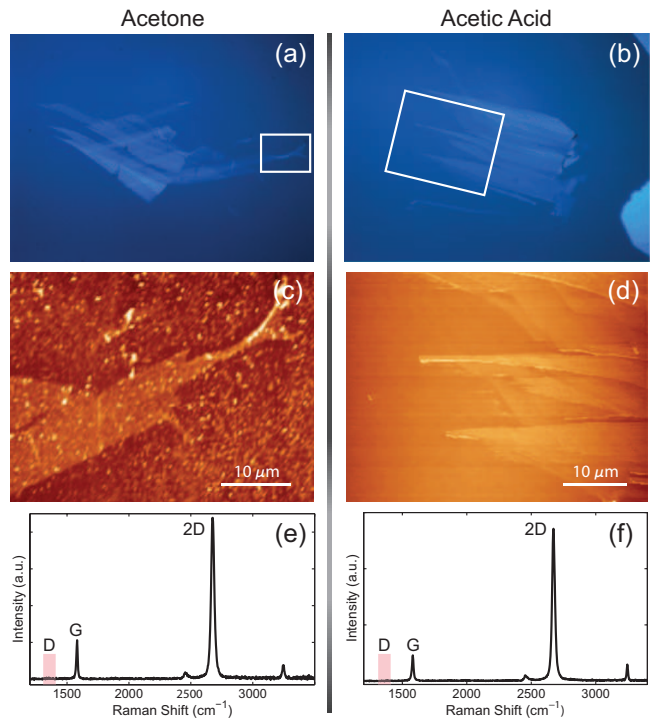}%
\caption{Comparison of graphene transfer methods. (a,c,d) Standard acetone based approach and (b,d,f)  acetic acid method.
The images show a graphene sample that has been transferred on a glass substrate.
(a,b)  Optical images viewed under a differential interference contrast (DIC) light microscope.  (c,d)  Topographic images recorded with AFM, and (e,f)  Raman spectra.  The red box indicates the D band frequency range.}
\label{acetone-acetic}
\end{center}
\end{figure}

%
Figure $\,$\ref{acetone-acetic} shows a comparison between acetone and acetic acid based transfer methods.  (a) and (b) are optical images recorded with DIC light microscopy, and (c) and (d) are topographic images recorded with AFM. (e) and (f) are representative Raman spectra that feature several characteristic phonon modes. For either transfer method we observed no folding or tearing of the graphene layer.\\[-1.2ex]

The Raman spectra were acquired by placing the samples on an inverted microscope.  The sample was excited with a 532\,nm laser through an air objective (NA = 0.7), which was also used for collection.  The signal was sent to either an avalanche photodiode (APD) or a combination of spectrometer and charge-coupled device (CCD).  The samples were raster scanned through the focus with an x-y piezo scan stage to form a confocal Raman 2D band image.  Using this confocal image, the laser was moved to specific locations on the graphene sample to obtain spectra.\\[-1.2ex]

Figure $\,$\ref{acetone-acetic}(e,f) show Raman spectra taken on an acetone and acetic acid cleaned sample, respectively.  The Raman spectra exhibit the first-order bond stretching G band centered at $\sim1580\,\rm{cm^{-1}}$, and the two-phonon 2D (or G$^\prime$) band centered at $\sim2700\,\rm{cm^{-1}}$.  Both spectra are characteristic of single layer graphene, as can be seen from the lineshape of the 2D band~\cite{ferrari06a,malard09}, and neither show additional peaks due to a chemical alteration of the graphene.  It is also important to note the absence of the disorder-induced D band [red box in Fig. $\,$\ref{acetone-acetic}(e-f)] ~\cite{casiraghi09b,cancado04,lucchese10} which indicates that neither transfer process damages the sp$^2$ bonds of the graphene flakes. \\[-1.2ex]

The residue from the acetone transfer method is not observable optically or spectroscopically, but is clearly visible in topography [Fig. $\,$\ref{acetone-acetic}(c)].  The topographic images were taken using a commercial AFM  operating in semi-contact mode.  The residue on this flake ranges in size from $\sim\,0.2-2.0\,\rm{\mu m}$.  By dissolving the PMMA in an acetic acid solution, we have been able to remove this residue from the graphene, as shown in Fig. $\,$\ref{acetone-acetic}(d).\\[-1.2ex]

To further illustrate the cleanliness of our acetic acid transfer method as well as demonstrate the effects of contaminants on the optical properties of a graphene flake, we acquired fluorescence quenching images of acetic acid and acetone cleaned samples.  The total decay rate of the excited state of a fluorescent molecule ($\gamma$) can be written as

\begin{equation}
\gamma = \gamma_{r} + \gamma_{nr}, 
\label{rate}
\end{equation}

where $\gamma_r$ and $\gamma_{nr}$ are the radiative and non-radiative rates, respectively.  As a fluorescent molecule is brought closer to the graphene surface, $\gamma$ increases due to the Purcell effect\cite{purcell46,novotny06b}.  However, the separation distance determines whether $\gamma_r$ or $\gamma_{nr}$ dominates.  In the case of a metal film, for distances larger than $\approx\,5\,\rm{nm}$, $\gamma_r$ dominates.  If separation distance is less than $5\,\rm{nm}$, then the decay process is primarily non-radiative ($\gamma_{nr} >> \gamma_r$), which is known as fluorescence quenching.  If there is a dielectric spacer between the molecule and the metal film then fluorescence is still observed and the thickness of the dielectric spacer determines the strength of the fluorescence signal.  Therefore, a clean graphene flake coated with fluorescent dye should show uniform quenching, whereas in the presence of contaminants on the surface, the dye molecules can still fluoresce.  Based on this, fluorescence quenching provides an optical probe for local surface contamination.\\[-1.2ex]

%
\begin{figure}[h!]
\begin{center}
\includegraphics[width = 8.5cm]{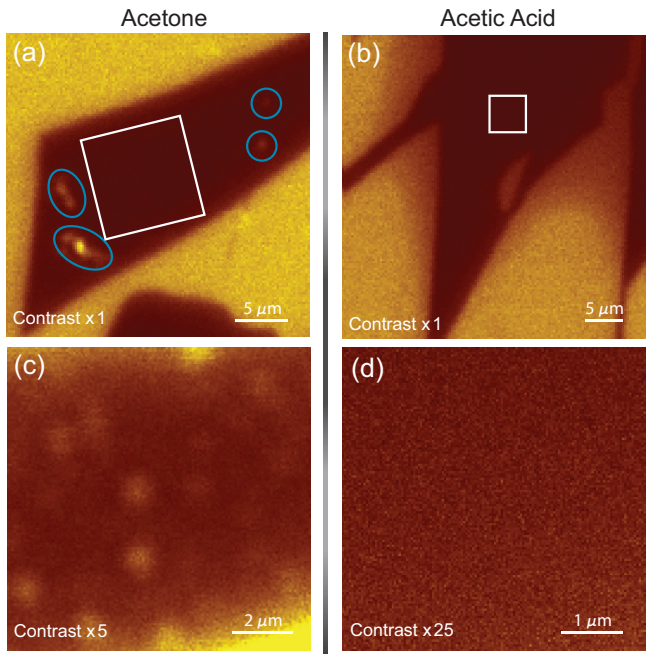}%
\caption{Confocal imaging showing quenching of fluorescent dye on areas where graphene is present.  (a)-(b) Post-transfer graphene samples using the acetone method and acetic acid method.  Bright spots (circled in blue) indicate local surface contaminants.  White boxes indicate the area where confocal images were taken for images (c)-(d).  (c) Local contaminants are visible on the graphene sample even though they are not obvious in (a).  To further illustrate the cleanliness of the acetic acid method, a similar scan (d) was performed for the graphene sample shown in (b).}
\label{fluor_quench}
\end{center}
\end{figure}
%
To visualize surface contaminants with fluorescence quenching we spin-coated Cy3 molecules on a sample prepared by the acetone method, and on another sample prepared by acidic acid procedure. The dye concentration and the spin-coating conditions were adjusted such that a monolayer of Cy3 is formed on the sample surface. Fluorescence images were recorded using the same optical microscope described above.  Figure\,\ref{fluor_quench}(a) shows the results for the acetone sample.  Local contaminates are clearly seen and are circled in blue.  Figure\,\ref{fluor_quench}(c) shows a zoomed-in image of the white box in (a), where additional weaker fluorescent spots, that are not visible in (a), can be observed.  Figures\,\ref{fluor_quench}(b) and (d) show fluorescence quenching images for the acetic acid treated sample, demonstrating the topographic cleanliness of our transfer procedure.  It is important to note that the contrast in Fig.\,\ref{fluor_quench}(d) has been scaled by $5\times$ compared to Fig.\,\ref{fluor_quench}(c) because the residual background fluorescence signal is considerably lower. The data shows that Cy3 molecules are evenly deposited on the acetic-acid treated graphene surface and that their fluorescence is almost entirely quenched. \\[-1.2ex]

In summary, we have demonstrated a simple graphene transfer method that yields very clean graphene surfaces. The method eliminates most of the residual contaminants that are present in the standard acetone-based transfer method.  Our technique yields defect-free graphene surfaces that can be deposited on a variety of substrates, allowing for the fabrication of more sophisticated graphene devices where stacking would be required.

\begin{acknowledgments}
The authors would like to thank Luiz Gustavo Can\c{c}ado and Palash Bharadwaj for fruitful discussions.
This research was funded by the U.S. Department of Energy (DE-FG02-05ER46207).\\[3ex]
\end{acknowledgments}


\end{document}